\documentclass[pra, twocolumn, superscriptaddress]{revtex4-2}
\setcitestyle{numbers,square}
\usepackage{graphicx}
\usepackage{dcolumn}
\usepackage{bm}

\usepackage[utf8]{inputenc}
\usepackage[T1]{fontenc}
\usepackage{mathptmx}
\usepackage{etoolbox}
\usepackage{natbib}

\makeatletter
\def\@email#1#2{%
 \endgroup
 \patchcmd{\titleblock@produce}
  {\frontmatter@RRAPformat}
  {\frontmatter@RRAPformat{\produce@RRAP{*#1\href{mailto:#2}{#2}}}\frontmatter@RRAPformat}
  {}{}
}%
\makeatother
\begin{document}

\title[A Semiconductor Photon Bose-Einstein Condensate as a Practical Light Source for Ranging Finding]{A Semiconductor Photon Bose-Einstein Condensate as a Practical Light Source for Ranging Finding}

\author{Ross C. Schofield}
  \email{ross.schofield@imperial.ac.uk}

\author{Daniel Lim}%
\author{Nathan R. Gemmell}
\affiliation{ 
Department of Physics, Blackett Laboratory, Imperial College London, London
}%
\author{Edmund Clarke}
\author{Ian Farrer}
\author{Aristotelis Trapalis}
\author{Jon Heffernan}
\affiliation{%
EPSRC National Epitaxy Facility, University of Sheffield, Sheffield
}%
\author {Rupert F. Oulton}
\affiliation{ 
Department of Physics, Blackett Laboratory, Imperial College London, London
}%

\date{\today}

\begin{abstract}
Here we report the measurement of thermal photon statistics from a semiconductor photon Bose-Einstein condensate operating just above the condensation threshold. We identify a regime where coherent, single mode emission occurs while still demonstrating significant photon bunching. Taking advantage of the photon bunching, along with the continuous-wave operation and high photon flux, we demonstrate optical range sensing using a photon Bose-Einstein condensate. We characterise the precision of the range measurement and analyse the dependence on the condensate's pump power and resulting coherence properties. 

\end{abstract}

\maketitle

\section{Introduction}
\noindent Condensates of light have often been suggested as novel light sources \cite{Bloch2022} with high spatial coherence \cite{Marelic2016b}, low threshold \cite{Byrnes2014}, and thermal photon statistics \cite{Schmitt2014,Emreztrk2023}. However, limitations in operation mean they have not been practical. Exciton-polariton condensates typically operate at cryogenic temperatures in order to stabilize the low binding energy of excitons in inorganic semiconductors \cite{Byrnes2014} whilst condensates based on organic dyes are susceptible to triplet state shelving and degradation if not excited using a pulsed laser \cite{Klaers2010Condensation}. The recent experimental realization of semiconductor-based room temperature photon Bose-Einstein condensates (BECs) now points to a technologically viable system, capable of continuous wave (cw) operation with reduced operating complexity, increased light output, and longer device lifetime \cite{Schofield2024}.

In this work we use the thermal statistics of a semiconductor photon BEC operated just above condensation threshold to perform range sensing. Unlike conventional optical ranging techniques, which assess photon time-of-flight directly \cite{Bender1973,Pellegrini2000}, here we use the Hanbury Brown–Twiss (HBT) effect to determine the time delay between two detectors \cite{Brown1956}. This requires a thermal or pseudo-thermal "bunched" light source, where the peak in second-order intensity correlations reveals the time delay and distance \cite{Zhu2012}. This approach benefits from being less sensitive to phase noise and can work with incoherent or scattered light, for example through turbulence, and background noise. Unfortunately, the coherence time of most thermal light sources, such as LEDs, is too short to measure (sub ps) using an HBT interferometer. Ranging with thermal light from LEDs has been possible with spectral filtering to achieve detectable coherence times \cite{Tan2023}, but this requires precise filtering and can lead to poor system efficiency. Operating such a system just below laser threshold ensures sufficient light emission without the instability of laser emission near threshold. Alternatively, pseudo-thermal lasers, produced when coherent laser light passes a rotating ground glass diffuser \cite{Zhu2012,Clark2024}, are efficient and bright, but have limited precision due to long ($\sim\mu$s) coherence times. The stable, single-mode, and coherent emission with thermal statistics from a photon BEC at threshold offers a compromise on both efficiency and precision with reduced system complexity. Here we use the light produced directly and without modification for range sensing.

\begin{figure}
    \centering
    \includegraphics[width=0.8\linewidth]{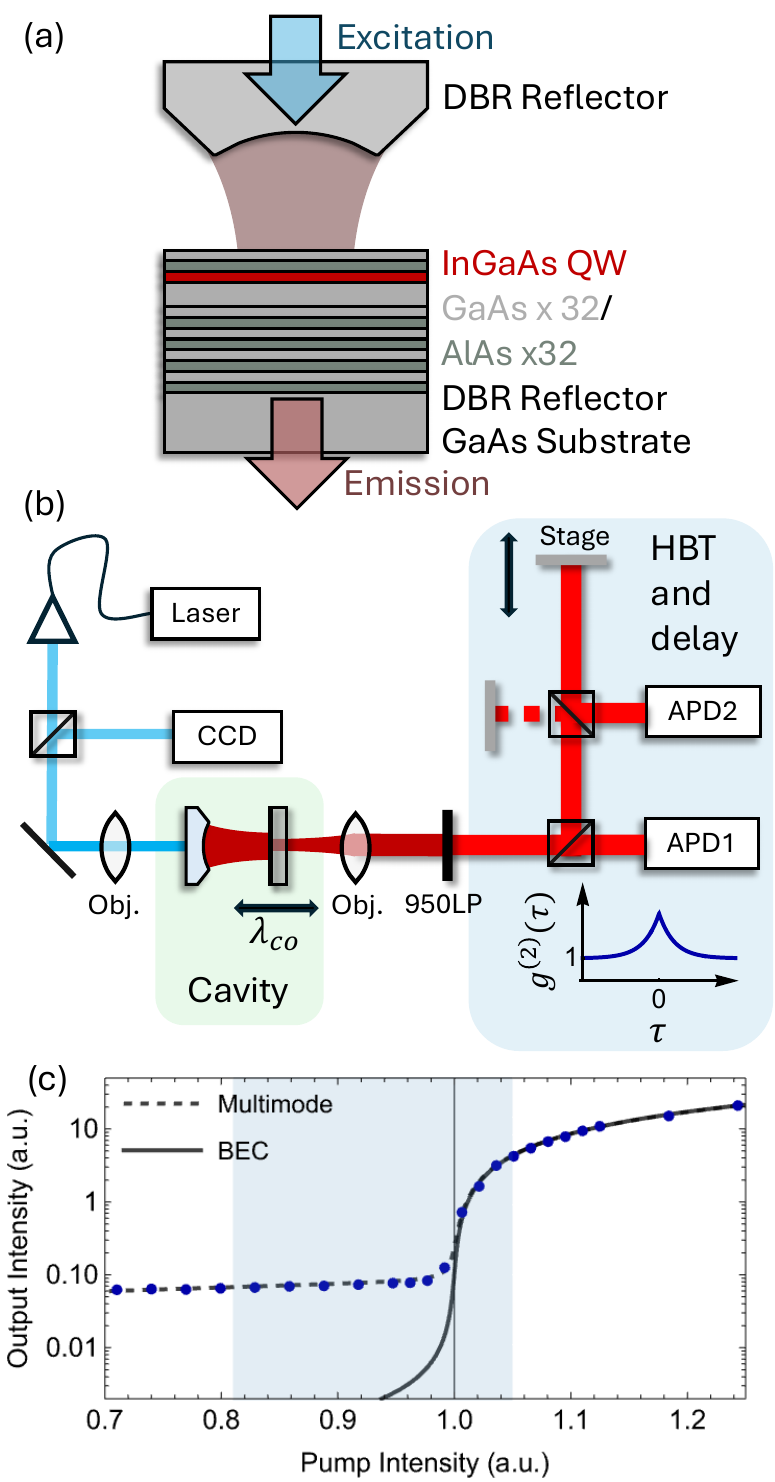}
    \caption{Cavity and experimental setup. (a) Schematic diagram of the open semiconductor microcavity. (b) Schematic diagram of the experimental setup. A fibre-coupled laser is used to excite the semiconductor in the cavity through an objective lens (Obj.). A CCD camera (CCD) is used to monitor cavity length by imaging an interference pattern from a broadband LED. Emission (red line) is collimated by an objective lens (Obj.), sent through a 950~nm LP filter (950LP) and sent to the HBT interferometer, highlighted in the blue box. The first APD (APD1) starts the coincidence timer, and the second APD (APD2) stops the timer. The second APD is placed after a Michelson interferometer with a variable delay stage (STAGE). The second arm, indicated with a dashed line, is blocked unless otherwise stated. (c) Input-output curve for the photon BEC. Data (points) are from power measurements fit with a multimode rate equation model (dashed line), shown with the single mode component of the model (solid line) \cite{Schofield2024}. The vertical line is $P=P_c$ and shaded region represents region of interest for this work. }
    \label{fig:fig1}
\end{figure}

We begin by detailing the cavity structure and experimental setup, and characterize the thermal character of the photon BEC emission through the second-order coherence function, $g^{(2)}(\tau)$. Next, we demonstrate the basic ranging principle by using the peak in the $g^{(2)}(\tau)$ measurement to determine relative time of flight, and distance, between the two detectors of the HBT interferometer. We also show the system's capability for detecting two objects at different positions in a single measurement. We finish by assessing how the measurement and system parameters of the photon BEC affect the precision and identify the optimal parameter regime. 

\section{Experimental Setup}

The open semiconductor microcavity that we use to create the BEC is shown in Fig.~1(a). Full details of BEC vs laser operation and sample fabrication are detailed in Ref.~\cite{Schofield2024}. The cavity is formed by a spherical mirror and a planar half-cavity, with the latter comprising a mirror and a single InGaAs quantum well. The curved mirror is a commercial half-inch Bragg mirror (Layertec) with a radius of curvature of 0.1~m and reflectivity of $\geq 99.99\%$, and has been ground to a diameter of 1~mm. The planar half-cavity consists of a 32-repeat GaAs/AlAs DBR, with a reflectivity $\geq 99.95\%$, on top of which there is a single InGaAs quantum well, with a peak absorption of $1\%$ at the quantum confined bandgap near 925~nm. 

For all measurements in this work the cavity is operating with the 14th longitudinal mode tuned to 950~nm. The cavity length is maintained using a piezoelectric crystal tuned using feedback from optical interference fringes from illumination with a 830 nm laser diode, imaged on a camera (CCD).

We excite the photon BEC using a 785~nm laser through the pass-band of the curved mirror. The laser is focused using a NIR objective lens (20X Plan Apochromat, Mitutoyo). This is primarily absorbed by the bulk GaAs with generated electron-hole pairs migrating to the InGaAs quantum well, where they recombine to emit photons into the cavity mode near 950 nm. Electrical injection of photon BECs is also possible with monolithic vertical-cavity surface emitting laser type devices, but requires careful tuning of system parameters as discussed in Ref.~\cite{Pieczarka2024}. Repeated absorption and emission events within the cavity allow the light to reach thermal equilibrium \cite{Figueiredo2025,Pelous2023,Wyborski2025}. As the pump intensity increases so does the photon density inside the cavity. Above the critical density a phase change occurs from a thermalised photon gas to a Bose-Einstein condensate, marked by a massive coherent occupation of the lowest energy state of the system, the fundamental mode of the cavity \cite{Klaers2010Condensation}. Figure~1(c) shows the unfiltered input-output light curve for this device, fit with a multimode laser model \cite{Schofield2024}, clearly showing the increase in output intensity upon condensation. The single mode component of the model is plotted to show the change in population of the BEC.

The optical setup used for measurements is shown in Fig.~1(b). The emission is collimated by an objective lens (10X Plan Apochromat, Mitutoyo) and sent through a 950 nm long-pass filter to reduce background spontaneous emission reaching the detectors. The emission then passes through a modified Hanbury Brown-Twiss interferometer to two avalanche photodiodes (APDs - ID100, ID Quantique). APD1 is placed immediately after a 10:90 beamsplitter while APD2 is after a Michelson interferometer with a variable delay stage (DDS600, Thorlabs). All measurements except those in Section 4B have one arm of the interferometer blocked. Both detectors have 900 nm long-pass filters to reduce detection events from room light. A timing card (HydraHarp 400, Picoquant) is used to record time intervals between subsequent detection events on APD 1 and 2, which are then displayed as a histogram. Range finding measurements use ND filters to reduce APD count rates to $<10^6$ s$^{-1}$, with photon statistics collected over 30 seconds.

\section{Second-Order Coherence Function Measurement}
\begin{figure*}
    \centering
    \includegraphics[width=0.8\linewidth]{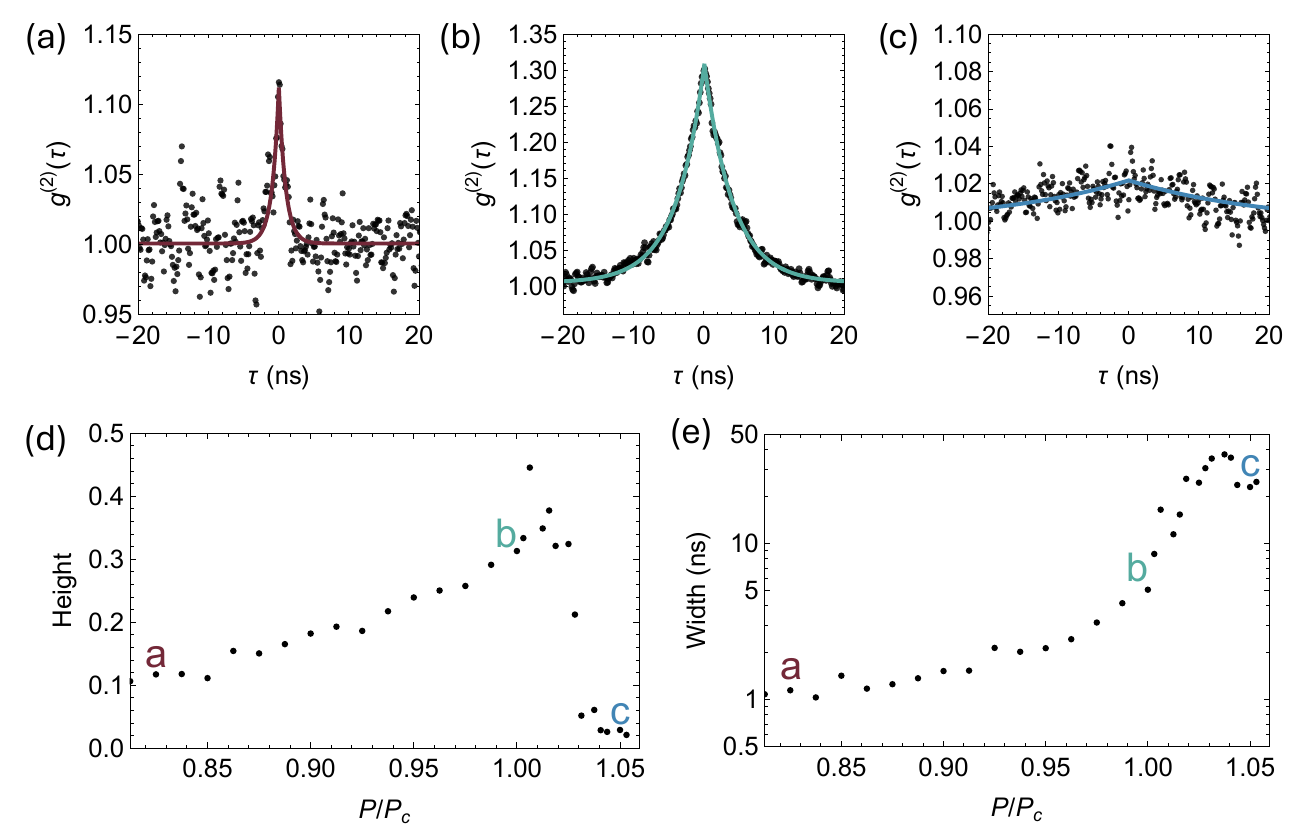}
    \caption{The height and width of the second-order correlation function $g^{(2)}(\tau)$ as a function of pump intensity. (a-c) $g^{(2)}(\tau)$ measurements at $P/P_c$ = 0.83, 1, 1.05 respectively. Data (points) are fit with the exponential decay shown in Eq.~1 (line). Labels in (d) and (e) indicate these measurements. (d) Height of the $g^{(2)}(\tau)$ as a function of power, where the Poissonian baseline value of 1 is subtracted. (e) Width of the $g^{(2)}(\tau)$ as a function of power. }
    \label{fig:enter-label}
\end{figure*}

Typically a coherent light source has Poissonian intensity fluctuations and therefore constant $g^{(2)}(\tau)=1$. A typical thermal light source instead shows bunching and $g^{(2)}(\tau=0)=2$. Condensates of light, despite being coherent light sources, also demonstrate thermal statistics at low condensate fractions due to their relatively large excitation reservoirs \cite{Schmitt2018}, with a transition from thermal to Poissonian statistics occurring only after full condensate formation. This has been widely studied in exciton-polariton condensates \cite{Kim2016} and dye-based photon condensates \cite{Emreztrk2023}, and here we present the first study of this in a semiconductor-based photon BEC. 

Figures~2(a), (b), (c) show the $g^{(2)}(\tau)$ measurement at increasing pump powers $P=0.83P_c$, $P_c$, and $1.05 P_c$ respectively. $P_c$ is the critical pump power, or pump power with the greatest increase in output intensity, as shown by the vertical line in Fig.~1(c). The data are fit with 
\begin{equation}
    g^{(2)}(\tau) = 1 + a e^{-|\tau-\tau_0|/\tau_l},\label{eqn:eqn1}
\end{equation}
where $a$ is a bunching amplitude, $\tau_0$ is the time delay between detectors, and $\tau_l$ is the width of the bunching peak. The height can be expressed as
\begin{equation}
   a = \frac{\bar n^2}{(\bar n + b)^2}\left(\frac{\Delta n}{\bar n^2}-1\right)
\end{equation}
where $\bar n$ is the mean photon number of the BEC mode, $b$ represents all other detection events (background, dark counts), and $\Delta n$ is the fluctuation in photon number of the BEC mode. The first term characterises the reduction due to uncorrelated detections and the second represents the inherent bunching of the BEC light \cite{Kim2016}. $\Delta n$ can be related to the effective reservoir size of the condensate and depends on the number of excitations in the system, the carrier density, and the spectral overlap of the BEC mode with the thermalisation medium, in this case the quantum well \cite{Schmitt2014}.

Whilst $g^{(2)}(\tau=0)$ should equal 2 before condensation occurs, we measure a significantly lower value of 1.11. The 950~nm long pass filter does not filter all emission from higher order transverse modes, meaning $b$ is large relative to  $\bar n$. As we increase the pump power $g^{(2)}(\tau=0)$ initially increases, as $\bar n$ increases and $b$ is effectively constant. This can be seen in Fig.~2(b) and (c), and the trend in Fig.~2(d). However, as pump power and $\bar n$ further increases, gain clamping occurs and the system enters the canonical regime. Eventually $ \bar n^2\gg\Delta n$, and Poissonian intensity fluctuations are reached with $g^{(2)}(\tau=0)\to 1$, as shown in Fig.~2(e), and we limit our analysis to before this point. The elevated $g^{(2)}(0)>1$ region for range measurements thus extends over $1.05>P/P_c>0.8$. The maximum height of $g^{(2)}(0)$ could be increased by further filtering, spectral and polarisation, to isolate the ground state mode of the cavity. This is, however, unnecessary for the use case highlighted here, and we opt for a experimentally simple implementation. 

Additionally, we see the $g^{(2)}(\tau)$ function initially broadens with increasing power, before plateauing and then subsequently falling as shown in Fig.~2(e). If we assume a modified Siegert relationship holds for this region, 
\begin{equation}
    g^{(2)}(\tau) = 1 + a|g^{(1)}(\tau)|^2, \label{eqn:eqn3}
\end{equation}
where $g^{(1)}(\tau)$ is the first-order coherence function, the broadening is due to the increasing coherence of the condensate above the critical intensity. We note studies that show this modified Siegert relationship holds near threshold for $g^{(2)}(0)>1.1$ for devices in the weak coupling regime \cite{Drechsler2022,Ulrich2007}. Using Eqn.~\ref{eqn:eqn3} we see that the coherence time is twice the width of the bunching feature, reaching $>70$ ns as the condensate forms. Further investigation is required to determine if the coherence begins to fall after this point, in-line with both theoretical and experimental studies on photon BECs and exciton-polariton condensates \cite{Kim2016,Tang2023a}, or if instead the Siegert relationship no longer holds at higher photon numbers \cite{Schmitt2018}.

\section{Range Finding}
\subsection{Experimental Demonstration}
\begin{figure*}
    \centering
    \includegraphics[width=0.8\linewidth]{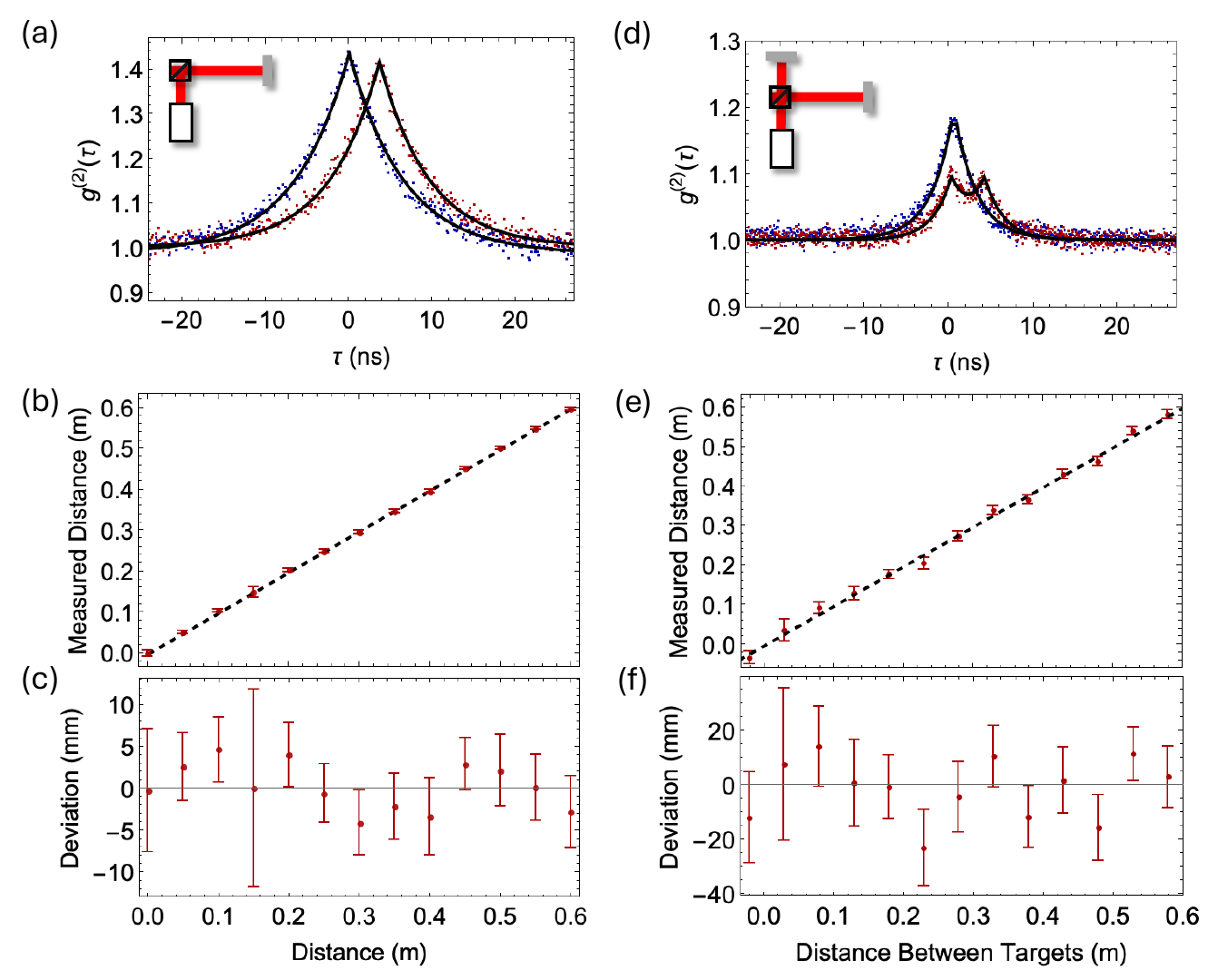}
    \caption{Distance measurement using thermal light. (a) Data (points) from two $g^{(2)}(\tau)$ measurements differing by approximately 1 m of optical delay, showing a clear shift in bunching feature. Data are fit with Eq.~1 (lines). (b) Measured distance (points) as a function of optical path delay. Dashed line is $x=y$. (c) Deviation from actual distance (points) as a function of distance. (d) Data (points) from two $g^{(2)}(\tau)$ measurement where one arm of the HBT interferometer has two paths differing by approximately 0.1 m (blue) and 1 m (red) of optical delay. The 0.1 m measurement looks like a single broadened peak, and the 1 m measurement shows two clear bunching features. Data are fit with Eq.~1 (line), modified to have two peaks. (e) Measured distance (points) as a function of actual distance, expressed as distance between objects. Dashed line is $x=y$. (f) Deviation from actual distance (points) as a function of distance.}
    \label{fig:enter-label}
\end{figure*}

We use the bunching feature in $g^{(2)}(\tau)$ to determine the relative path length differences between two detectors. Figure~3(a) shows two $g^{(2)}(\tau)$ measurements taken with approximately 1~m of additional optical delay in one of the arms, showing a clear shift in the bunching peak. These measurements were taken with $P\approx P_c$, where $g^{(2)}(0)\approx 1.3$. The central location of the $g^{(2)}(\tau)$ bunching peak can be tracked through fitting with Eqn.~\ref{eqn:eqn1}, where the shift in the centre of the feature is determined by $\tau_0$, the delay between the detectors. Accounting for the zero delay offset the time delay reveals the distance between detectors.

To test the effectiveness of this ranging approach we used a translation stage to shift a mirror on a delay stage up to 0.6~m, corresponding to 1.2~m of optical delay, and use Eqn.~\ref{eqn:eqn1} to determine the distance at each mirror position. Figure~3(b) shows the correspondence between actual and measured mirror positions, which works well over the full delay range. The deviation is shown in Fig.~3(c), with all measurements falling within 5~mm of the actual delay. The magnitude of the error is independent of the measured distance.

\subsection{Multi-Target Ranging}
As a proof of principle measurement we demonstrate how this technique can measure relative distance between two objects simultaneously; in effect, depth imaging. We unblock the second arm on the Michelson interferometer and send the BEC emission to two mirrors. A $g^{(2)}(\tau)$ measurement is then performed, with an example shown in Fig.~3(d) for long time delay. Two bunching features are clearly visible, corresponding to the two possible paths. We note for this measurement we reduced power to $P\approx0.92P_c$ to allow the two peaks to be better visually distinguished.

The $g^{(2)}(\tau)$ features can then be fit in order to extract the difference in time-of-flight for the two paths, and therefore relative distance. Figure~3(e) shows the actual and measured distance between the two mirrors, whilst Fig.~3(f) shows the deviation. The results are generally similar to those shown in Fig.~3(b-c) for a single distance measurement, albeit with larger errors due to fitting two bunching features and lower count rates. Notably, if the width of the bunching feature is well known, distances of $\approx 2$~cm can be measured, corresponding to flight times shorter than the width. This method does not give access to any spatial information, however this can be achieved by either using an APD array or raster scanning the imaging beam \cite{Wagner2021}. Combining a spatial method with this multi-target range finding should increase the effective resolution of the image, beneficial when considering the difficulties in achieving high spatial resolution conventionally.

\subsection{Measurement Precision}

We now move onto looking at how system and measurement parameters affect the precision and uncertainty of distance measurement. We begin by looking at the relationship between measured distance, integration time and uncertainty in Fig.~4(a). We compare the error magnitude for no (blue) and 0.6~m (red) distances and see no significant difference. This matches with the expected behaviour from laser range finding, including that using bunched light \cite{Zhu2012,Tan2023}. The constant uncertainty means optical range finding is much more relatively precise for long distance measurement. 

For our experimental count rates, we find that integration times beyond 80 seconds offer diminishing returns in measurement precision, and depending on desired accuracy, integration for less than 20 seconds can achieve promising results. Whilst not representative of real world conditions as we use a mirror, not a likely much less reflective object, the Michelson interferometer with a blocked arm does already reduce count rates by a factor of 4.

\begin{figure}
    \centering
    \includegraphics[width=0.8\linewidth]{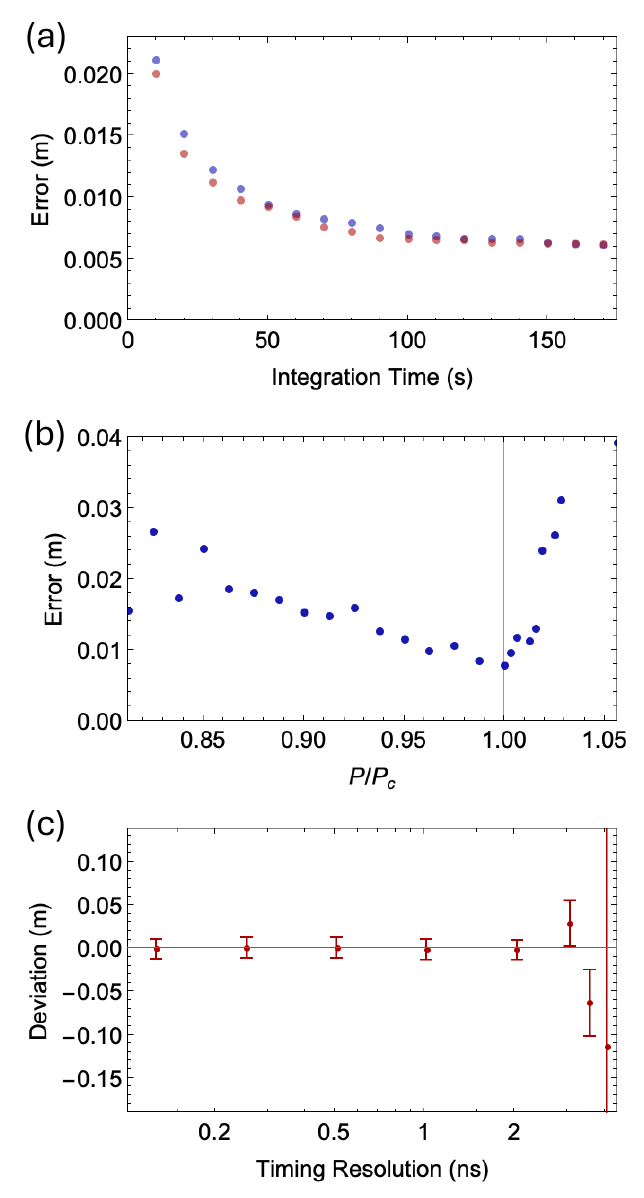}
    \caption{Dependence of error in distance measurement as a function of system parameters. (a) Error magnitude as a function of integration time (points) for no (blue) and 0.6 m (red) delay for $P= P_c$. Both tend to the same value showing independence of measurement error on distance. (b) Error magnitude as a function of $P/P_c$. (c) Average deviation and error magnitude for distance measurement as a function of timing resolution for $0.9 <P/P_c\leq 1$. Timing resolution is reduced through re-binning data.}
    \label{fig:placeholder}
\end{figure}

As the height and width of the $g^{(2)}(\tau)$ measurement are directly power dependent, we now investigate the optimal regime for maximising precision in the measurement. The precision in the $\tau_0$ value for the data shown in Fig.~2 is shown in Fig.~4(b). We initially see that the error decrease with pump power, likely due to the increase in count rate over the integration period, before rapidly increasing again. The dashed line on the plot indicates the power where the width of the bunching feature begins to rapidly increase, and the solid line is the power where the $g^{(2)}(\tau=0)$ height plateaus. Fewer coincidence counts contributing to a broader feature explains this decrease in precision. There is a significant power range where the error is reasonably low, giving a good tolerance to variance in pump power and cavity length, which affects $P_c$~\cite{Schofield2024}.

The APDs and timing card combination used have timing resolution on the order of 100~ps, however the width of the $g^{(2)}(\tau)$ feature is significantly broader than this. In Fig.~4(c) we investigate the relationship between timing resolution and the accuracy and precision of the distance measurement. We take the mean value of measurements for $0.9 <P/P_c< 1$, which have a width of approximately 2.2~ns. The resolution is changed by binning the original data taken with 128~ps timing resolution. There is no significant reduction in precision until the timing resolution exceeds the bunching feature width. Using higher pump powers to measure a broader bunching feature could further relax the timing precision required, however this is limited by the rapidly falling $g^{(2)}(\tau=0)$ value. This could be addressed by moving to using a shorter cavity with better overlap between the fundamental mode and the quantum well emission and absorption spectra, which would allow for thermal fluctuations to be seen at higher powers \cite{Schmitt2014}.

Finally, we consider the brightness and spectral density of our system~\cite{Tan2023}. When optimised for precision the total photon flux is in the region of $5*10^{12}$~s$^{-1}$. Spectral measurements give a spectrometer-limited linewidth value of $0.05$~nm, however if we use the modified Siegert relation this would give a linewidth of $0.1$~pm. This gives a spectral density of $1*10^{14}$ to $5*10^{16}$ photons/second/nm. If we instead operate at the maximum coherence value, values of $4*10^{17}$ photons/second/nm can be achieved. We note that there is further room to improve the count rate of our device to achieve even higher values.

\section{Conclusion}

In conclusion, we have demonstrated the first application of a room-temperature, continuous-wave semiconductor photon Bose-Einstein condensate as a light source for time-of-flight range sensing. By characterising the above-threshold thermal photon statistics, we established that the second-order coherence bunching peak, $g^{(2)}(\tau)$, provides a robust feature for determining optical delay. We measure distances up to 0.6 m with a precision of approximately 5 mm, and critically, we identified the optimal pump power regime where measurement uncertainty is minimised by balancing photon flux against the coherence characteristics of the condensate. 

Beyond the specific application of range sensing, this work serves as a broader demonstration that continuous-wave, room-temperature semiconductor photon BECs have now achieved the brightness and stability required to be considered practical light sources. This achievement bridges the gap between fundamental studies of macroscopic quantum states and practical metrology. While photon BECs have previously been explored for their unique properties, our results establish a clear pathway towards their use in real-world applications such as LiDAR or depth imaging. The ability to harness thermal photon bunching for ranging offers a new approach for coherent sensing systems. Furthermore, our proof-of-principle demonstration of simultaneous multi-distance measurement highlights the potential for this method to be extended to full 3D imaging.

We acknowledge that the current demonstration has limitations. The overall photon flux, while high for a condensate system, remains modest for long-range applications, and the experiments were performed using a high-reflectivity mirror rather than a diffuse, non-cooperative target. Future work should therefore focus on several key areas. Optimising the microcavity design by increasing the quantum well overlap with the cavity mode and operating the device nearer the quantum well bandgap as suggested by theory \cite{Schmitt2018} could significantly expand the operational power range where thermal statistics are prominent. Following this, testing the system's performance on various real-world surfaces and developing parallel detection schemes would be the logical next steps toward realising a complete, single-photon sensitive imaging system based on a photon BEC.

\begin{acknowledgments}
We thank J. Murphy for optical workshop support and S. Cussell for mechanical workshop support. This work was supported by the UK Engineering and Physical Sciences Research Council project ‘Near-equilibrium thermalised quantum light’ (EP/S000755/1) (R.C.S., E.C.,  I.F.,  J.H.,  R.F.O.).
\end{acknowledgments}

\section*{Data Availability Statement}

The data that support the findings of this study are available from the corresponding author upon reasonable request.

\bibliography{export}

\begin{thebibliography}{23}%
\makeatletter
\providecommand \@ifxundefined [1]{%
 \@ifx{#1\undefined}
}%
\providecommand \@ifnum [1]{%
 \ifnum #1\expandafter \@firstoftwo
 \else \expandafter \@secondoftwo
 \fi
}%
\providecommand \@ifx [1]{%
 \ifx #1\expandafter \@firstoftwo
 \else \expandafter \@secondoftwo
 \fi
}%
\providecommand \natexlab [1]{#1}%
\providecommand \enquote  [1]{``#1''}%
\providecommand \bibnamefont  [1]{#1}%
\providecommand \bibfnamefont [1]{#1}%
\providecommand \citenamefont [1]{#1}%
\providecommand \href@noop [0]{\@secondoftwo}%
\providecommand \href [0]{\begingroup \@sanitize@url \@href}%
\providecommand \@href[1]{\@@startlink{#1}\@@href}%
\providecommand \@@href[1]{\endgroup#1\@@endlink}%
\providecommand \@sanitize@url [0]{\catcode `\\12\catcode `\$12\catcode `\&12\catcode `\#12\catcode `\^12\catcode `\_12\catcode `\%12\relax}%
\providecommand \@@startlink[1]{}%
\providecommand \@@endlink[0]{}%
\providecommand \url  [0]{\begingroup\@sanitize@url \@url }%
\providecommand \@url [1]{\endgroup\@href {#1}{\urlprefix }}%
\providecommand \urlprefix  [0]{URL }%
\providecommand \Eprint [0]{\href }%
\providecommand \doibase [0]{https://doi.org/}%
\providecommand \selectlanguage [0]{\@gobble}%
\providecommand \bibinfo  [0]{\@secondoftwo}%
\providecommand \bibfield  [0]{\@secondoftwo}%
\providecommand \translation [1]{[#1]}%
\providecommand \BibitemOpen [0]{}%
\providecommand \bibitemStop [0]{}%
\providecommand \bibitemNoStop [0]{.\EOS\space}%
\providecommand \EOS [0]{\spacefactor3000\relax}%
\providecommand \BibitemShut  [1]{\csname bibitem#1\endcsname}%
\let\auto@bib@innerbib\@empty
\bibitem [{\citenamefont {Bloch}\ \emph {et~al.}(2022)\citenamefont {Bloch}, \citenamefont {Carusotto},\ and\ \citenamefont {Wouters}}]{Bloch2022}%
  \BibitemOpen
  \bibfield  {author} {\bibinfo {author} {\bibfnamefont {J.}~\bibnamefont {Bloch}}, \bibinfo {author} {\bibfnamefont {I.}~\bibnamefont {Carusotto}},\ and\ \bibinfo {author} {\bibfnamefont {M.}~\bibnamefont {Wouters}},\ }\bibfield  {title} {\bibinfo {title} {Non-equilibrium bose–einstein condensation in photonic systems},\ }\href {https://doi.org/10.1038/s42254-022-00464-0} {\bibfield  {journal} {\bibinfo  {journal} {Nature Reviews Physics}\ }\textbf {\bibinfo {volume} {4}},\ \bibinfo {pages} {470} (\bibinfo {year} {2022})}\BibitemShut {NoStop}%
\bibitem [{\citenamefont {Marelic}\ \emph {et~al.}(2016)\citenamefont {Marelic}, \citenamefont {Zajiczek}, \citenamefont {Hesten}, \citenamefont {Leung}, \citenamefont {Ong}, \citenamefont {Mintert},\ and\ \citenamefont {Nyman}}]{Marelic2016b}%
  \BibitemOpen
  \bibfield  {author} {\bibinfo {author} {\bibfnamefont {J.}~\bibnamefont {Marelic}}, \bibinfo {author} {\bibfnamefont {L.~F.}\ \bibnamefont {Zajiczek}}, \bibinfo {author} {\bibfnamefont {H.~J.}\ \bibnamefont {Hesten}}, \bibinfo {author} {\bibfnamefont {K.~H.}\ \bibnamefont {Leung}}, \bibinfo {author} {\bibfnamefont {E.~Y.~X.}\ \bibnamefont {Ong}}, \bibinfo {author} {\bibfnamefont {F.}~\bibnamefont {Mintert}},\ and\ \bibinfo {author} {\bibfnamefont {R.~A.}\ \bibnamefont {Nyman}},\ }\bibfield  {title} {\bibinfo {title} {Spatiotemporal coherence of non-equilibrium multimode photon condensates},\ }\href {https://doi.org/10.1088/1367-2630/18/10/103012} {\bibfield  {journal} {\bibinfo  {journal} {New Journal of Physics}\ }\textbf {\bibinfo {volume} {18}},\ \bibinfo {pages} {103012} (\bibinfo {year} {2016})}\BibitemShut {NoStop}%
\bibitem [{\citenamefont {Byrnes}\ \emph {et~al.}(2014)\citenamefont {Byrnes}, \citenamefont {Kim},\ and\ \citenamefont {Yamamoto}}]{Byrnes2014}%
  \BibitemOpen
  \bibfield  {author} {\bibinfo {author} {\bibfnamefont {T.}~\bibnamefont {Byrnes}}, \bibinfo {author} {\bibfnamefont {N.~Y.}\ \bibnamefont {Kim}},\ and\ \bibinfo {author} {\bibfnamefont {Y.}~\bibnamefont {Yamamoto}},\ }\bibfield  {title} {\bibinfo {title} {Exciton-polariton condensates},\ }\href {https://doi.org/10.1038/NPHYS3143;SUBJMETA} {\bibfield  {journal} {\bibinfo  {journal} {Nature Physics}\ }\textbf {\bibinfo {volume} {10}},\ \bibinfo {pages} {803} (\bibinfo {year} {2014})}\BibitemShut {NoStop}%
\bibitem [{\citenamefont {Schmitt}\ \emph {et~al.}(2014)\citenamefont {Schmitt}, \citenamefont {Damm}, \citenamefont {Dung}, \citenamefont {Vewinger}, \citenamefont {Klaers},\ and\ \citenamefont {Weitz}}]{Schmitt2014}%
  \BibitemOpen
  \bibfield  {author} {\bibinfo {author} {\bibfnamefont {J.}~\bibnamefont {Schmitt}}, \bibinfo {author} {\bibfnamefont {T.}~\bibnamefont {Damm}}, \bibinfo {author} {\bibfnamefont {D.}~\bibnamefont {Dung}}, \bibinfo {author} {\bibfnamefont {F.}~\bibnamefont {Vewinger}}, \bibinfo {author} {\bibfnamefont {J.}~\bibnamefont {Klaers}},\ and\ \bibinfo {author} {\bibfnamefont {M.}~\bibnamefont {Weitz}},\ }\bibfield  {title} {\bibinfo {title} {Observation of grand-canonical number statistics in a photon bose-einstein condensate},\ }\href {https://doi.org/10.1103/PHYSREVLETT.112.030401/SUPPLEMENTAL.TEX} {\bibfield  {journal} {\bibinfo  {journal} {Physical Review Letters}\ }\textbf {\bibinfo {volume} {112}},\ \bibinfo {pages} {030401} (\bibinfo {year} {2014})}\BibitemShut {NoStop}%
\bibitem [{\citenamefont {Öztürk}\ \emph {et~al.}(2023)\citenamefont {Öztürk}, \citenamefont {Vewinger}, \citenamefont {Weitz},\ and\ \citenamefont {Schmitt}}]{Emreztrk2023}%
  \BibitemOpen
  \bibfield  {author} {\bibinfo {author} {\bibfnamefont {F.~E.}\ \bibnamefont {Öztürk}}, \bibinfo {author} {\bibfnamefont {F.}~\bibnamefont {Vewinger}}, \bibinfo {author} {\bibfnamefont {M.}~\bibnamefont {Weitz}},\ and\ \bibinfo {author} {\bibfnamefont {J.}~\bibnamefont {Schmitt}},\ }\bibfield  {title} {\bibinfo {title} {Fluctuation-dissipation relation for a bose-einstein condensate of photons},\ }\href {https://doi.org/10.1103/PhysRevLett.130.033602} {\bibfield  {journal} {\bibinfo  {journal} {Physical Review Letters}\ }\textbf {\bibinfo {volume} {130}},\ \bibinfo {pages} {33602} (\bibinfo {year} {2023})}\BibitemShut {NoStop}%
\bibitem [{\citenamefont {Klaers}\ \emph {et~al.}(2010)\citenamefont {Klaers}, \citenamefont {Schmitt}, \citenamefont {Vewinger},\ and\ \citenamefont {Weitz}}]{Klaers2010Condensation}%
  \BibitemOpen
  \bibfield  {author} {\bibinfo {author} {\bibfnamefont {J.}~\bibnamefont {Klaers}}, \bibinfo {author} {\bibfnamefont {J.}~\bibnamefont {Schmitt}}, \bibinfo {author} {\bibfnamefont {F.}~\bibnamefont {Vewinger}},\ and\ \bibinfo {author} {\bibfnamefont {M.}~\bibnamefont {Weitz}},\ }\bibfield  {title} {\bibinfo {title} {Bose–einstein condensation of photons in an optical microcavity},\ }\href {https://doi.org/10.1038/nature09567} {\bibfield  {journal} {\bibinfo  {journal} {Nature 2010 468:7323}\ }\textbf {\bibinfo {volume} {468}},\ \bibinfo {pages} {545} (\bibinfo {year} {2010})}\BibitemShut {NoStop}%
\bibitem [{\citenamefont {Schofield}\ \emph {et~al.}(2024)\citenamefont {Schofield}, \citenamefont {Fu}, \citenamefont {Clarke}, \citenamefont {Farrer}, \citenamefont {Trapalis}, \citenamefont {Dhar}, \citenamefont {Mukherjee}, \citenamefont {Millard}, \citenamefont {Heffernan}, \citenamefont {Mintert}, \citenamefont {Nyman},\ and\ \citenamefont {Oulton}}]{Schofield2024}%
  \BibitemOpen
  \bibfield  {author} {\bibinfo {author} {\bibfnamefont {R.~C.}\ \bibnamefont {Schofield}}, \bibinfo {author} {\bibfnamefont {M.}~\bibnamefont {Fu}}, \bibinfo {author} {\bibfnamefont {E.}~\bibnamefont {Clarke}}, \bibinfo {author} {\bibfnamefont {I.}~\bibnamefont {Farrer}}, \bibinfo {author} {\bibfnamefont {A.}~\bibnamefont {Trapalis}}, \bibinfo {author} {\bibfnamefont {H.~S.}\ \bibnamefont {Dhar}}, \bibinfo {author} {\bibfnamefont {R.}~\bibnamefont {Mukherjee}}, \bibinfo {author} {\bibfnamefont {T.~S.}\ \bibnamefont {Millard}}, \bibinfo {author} {\bibfnamefont {J.}~\bibnamefont {Heffernan}}, \bibinfo {author} {\bibfnamefont {F.}~\bibnamefont {Mintert}}, \bibinfo {author} {\bibfnamefont {R.~A.}\ \bibnamefont {Nyman}},\ and\ \bibinfo {author} {\bibfnamefont {R.~F.}\ \bibnamefont {Oulton}},\ }\bibfield  {title} {\bibinfo {title} {Bose–einstein condensation of light in a semiconductor quantum well microcavity},\ }\href {https://doi.org/10.1038/S41566-024-01491-2;SUBJMETA} {\bibfield  {journal} {\bibinfo
  {journal} {Nature Photonics}\ }\textbf {\bibinfo {volume} {18}},\ \bibinfo {pages} {1083} (\bibinfo {year} {2024})}\BibitemShut {NoStop}%
\bibitem [{\citenamefont {Bender}\ \emph {et~al.}(1973)\citenamefont {Bender}, \citenamefont {Currie}, \citenamefont {Dicke}, \citenamefont {Eckhardt}, \citenamefont {Faller}, \citenamefont {Kaula}, \citenamefont {Mulholland}, \citenamefont {Plotkin}, \citenamefont {Poultney}, \citenamefont {Silverberg}, \citenamefont {Wilkinson}, \citenamefont {Williams},\ and\ \citenamefont {Alley}}]{Bender1973}%
  \BibitemOpen
  \bibfield  {author} {\bibinfo {author} {\bibfnamefont {P.~L.}\ \bibnamefont {Bender}}, \bibinfo {author} {\bibfnamefont {D.~G.}\ \bibnamefont {Currie}}, \bibinfo {author} {\bibfnamefont {R.~H.}\ \bibnamefont {Dicke}}, \bibinfo {author} {\bibfnamefont {D.~H.}\ \bibnamefont {Eckhardt}}, \bibinfo {author} {\bibfnamefont {J.~E.}\ \bibnamefont {Faller}}, \bibinfo {author} {\bibfnamefont {W.~M.}\ \bibnamefont {Kaula}}, \bibinfo {author} {\bibfnamefont {J.~D.}\ \bibnamefont {Mulholland}}, \bibinfo {author} {\bibfnamefont {H.~H.}\ \bibnamefont {Plotkin}}, \bibinfo {author} {\bibfnamefont {S.~K.}\ \bibnamefont {Poultney}}, \bibinfo {author} {\bibfnamefont {E.~C.}\ \bibnamefont {Silverberg}}, \bibinfo {author} {\bibfnamefont {D.~T.}\ \bibnamefont {Wilkinson}}, \bibinfo {author} {\bibfnamefont {J.~G.}\ \bibnamefont {Williams}},\ and\ \bibinfo {author} {\bibfnamefont {C.~O.}\ \bibnamefont {Alley}},\ }\bibfield  {title} {\bibinfo {title} {The lunar laser ranging experiment},\ }\href
  {https://doi.org/10.1126/SCIENCE.182.4109.229} {\bibfield  {journal} {\bibinfo  {journal} {Science}\ }\textbf {\bibinfo {volume} {182}},\ \bibinfo {pages} {229} (\bibinfo {year} {1973})}\BibitemShut {NoStop}%
\bibitem [{\citenamefont {Pellegrini}\ \emph {et~al.}(2000)\citenamefont {Pellegrini}, \citenamefont {Buller}, \citenamefont {Smith}, \citenamefont {Wallace},\ and\ \citenamefont {Cova}}]{Pellegrini2000}%
  \BibitemOpen
  \bibfield  {author} {\bibinfo {author} {\bibfnamefont {S.}~\bibnamefont {Pellegrini}}, \bibinfo {author} {\bibfnamefont {G.~S.}\ \bibnamefont {Buller}}, \bibinfo {author} {\bibfnamefont {J.~M.}\ \bibnamefont {Smith}}, \bibinfo {author} {\bibfnamefont {A.~M.}\ \bibnamefont {Wallace}},\ and\ \bibinfo {author} {\bibfnamefont {S.}~\bibnamefont {Cova}},\ }\bibfield  {title} {\bibinfo {title} {Laser-based distance measurement using picosecond resolution time-correlated single-photon counting},\ }\href@noop {} {\bibfield  {journal} {\bibinfo  {journal} {Meas. Sci. Technol}\ }\textbf {\bibinfo {volume} {11}},\ \bibinfo {pages} {712} (\bibinfo {year} {2000})}\BibitemShut {NoStop}%
\bibitem [{\citenamefont {Brown}\ and\ \citenamefont {Twiss}(1956)}]{Brown1956}%
  \BibitemOpen
  \bibfield  {author} {\bibinfo {author} {\bibfnamefont {R.~H.}\ \bibnamefont {Brown}}\ and\ \bibinfo {author} {\bibfnamefont {R.~Q.}\ \bibnamefont {Twiss}},\ }\bibfield  {title} {\bibinfo {title} {Correlation between photons in two coherent beams of light},\ }\href {https://doi.org/10.1038/177027a0} {\bibfield  {journal} {\bibinfo  {journal} {Nature}\ }\textbf {\bibinfo {volume} {177}},\ \bibinfo {pages} {27} (\bibinfo {year} {1956})}\BibitemShut {NoStop}%
\bibitem [{\citenamefont {Zhu}\ \emph {et~al.}(2012)\citenamefont {Zhu}, \citenamefont {Chen}, \citenamefont {Huang},\ and\ \citenamefont {Zeng}}]{Zhu2012}%
  \BibitemOpen
  \bibfield  {author} {\bibinfo {author} {\bibfnamefont {J.}~\bibnamefont {Zhu}}, \bibinfo {author} {\bibfnamefont {X.}~\bibnamefont {Chen}}, \bibinfo {author} {\bibfnamefont {P.}~\bibnamefont {Huang}},\ and\ \bibinfo {author} {\bibfnamefont {G.}~\bibnamefont {Zeng}},\ }\bibfield  {title} {\bibinfo {title} {Thermal-light-based ranging using second-order coherence},\ }\href {https://doi.org/10.1364/AO.51.004885} {\bibfield  {journal} {\bibinfo  {journal} {Applied Optics, Vol. 51, Issue 20, pp. 4885-4890}\ }\textbf {\bibinfo {volume} {51}},\ \bibinfo {pages} {4885} (\bibinfo {year} {2012})}\BibitemShut {NoStop}%
\bibitem [{\citenamefont {Tan}\ \emph {et~al.}(2023)\citenamefont {Tan}, \citenamefont {Yeo}, \citenamefont {Leow}, \citenamefont {Shen},\ and\ \citenamefont {Kurtsiefer}}]{Tan2023}%
  \BibitemOpen
  \bibfield  {author} {\bibinfo {author} {\bibfnamefont {P.~K.}\ \bibnamefont {Tan}}, \bibinfo {author} {\bibfnamefont {X.~J.}\ \bibnamefont {Yeo}}, \bibinfo {author} {\bibfnamefont {A.~Z.~W.}\ \bibnamefont {Leow}}, \bibinfo {author} {\bibfnamefont {L.}~\bibnamefont {Shen}},\ and\ \bibinfo {author} {\bibfnamefont {C.}~\bibnamefont {Kurtsiefer}},\ }\bibfield  {title} {\bibinfo {title} {Practical range sensing with thermal light},\ }\href {https://doi.org/10.1103/PHYSREVAPPLIED.20.014060/FIGURES/5/MEDIUM} {\bibfield  {journal} {\bibinfo  {journal} {Physical Review Applied}\ }\textbf {\bibinfo {volume} {20}},\ \bibinfo {pages} {014060} (\bibinfo {year} {2023})}\BibitemShut {NoStop}%
\bibitem [{\citenamefont {Clark}\ \emph {et~al.}(2024)\citenamefont {Clark}, \citenamefont {Bishop}, \citenamefont {Cannon}, \citenamefont {Hadden}, \citenamefont {Dolan}, \citenamefont {Sinclair},\ and\ \citenamefont {Bennett}}]{Clark2024}%
  \BibitemOpen
  \bibfield  {author} {\bibinfo {author} {\bibfnamefont {R.~N.}\ \bibnamefont {Clark}}, \bibinfo {author} {\bibfnamefont {S.~G.}\ \bibnamefont {Bishop}}, \bibinfo {author} {\bibfnamefont {J.~K.}\ \bibnamefont {Cannon}}, \bibinfo {author} {\bibfnamefont {J.~P.}\ \bibnamefont {Hadden}}, \bibinfo {author} {\bibfnamefont {P.~R.}\ \bibnamefont {Dolan}}, \bibinfo {author} {\bibfnamefont {A.~G.}\ \bibnamefont {Sinclair}},\ and\ \bibinfo {author} {\bibfnamefont {A.~J.}\ \bibnamefont {Bennett}},\ }\bibfield  {title} {\bibinfo {title} {Measuring photon correlation using imperfect detectors},\ }\href {https://doi.org/10.1103/PhysRevApplied.22.064067} {\bibfield  {journal} {\bibinfo  {journal} {Physical Review Applied}\ }\textbf {\bibinfo {volume} {22}},\ \bibinfo {pages} {064067} (\bibinfo {year} {2024})}\BibitemShut {NoStop}%
\bibitem [{\citenamefont {Pieczarka}\ \emph {et~al.}(2024)\citenamefont {Pieczarka}, \citenamefont {Gębski}, \citenamefont {Piasecka}, \citenamefont {Lott}, \citenamefont {Pelster}, \citenamefont {Wasiak},\ and\ \citenamefont {Czyszanowski}}]{Pieczarka2024}%
  \BibitemOpen
  \bibfield  {author} {\bibinfo {author} {\bibfnamefont {M.}~\bibnamefont {Pieczarka}}, \bibinfo {author} {\bibfnamefont {M.}~\bibnamefont {Gębski}}, \bibinfo {author} {\bibfnamefont {A.~N.}\ \bibnamefont {Piasecka}}, \bibinfo {author} {\bibfnamefont {J.~A.}\ \bibnamefont {Lott}}, \bibinfo {author} {\bibfnamefont {A.}~\bibnamefont {Pelster}}, \bibinfo {author} {\bibfnamefont {M.}~\bibnamefont {Wasiak}},\ and\ \bibinfo {author} {\bibfnamefont {T.}~\bibnamefont {Czyszanowski}},\ }\bibfield  {title} {\bibinfo {title} {Bose–einstein condensation of photons in a vertical-cavity surface-emitting laser},\ }\href {https://doi.org/10.1038/s41566-024-01478-z} {\bibfield  {journal} {\bibinfo  {journal} {Nature Photonics}\ }\textbf {\bibinfo {volume} {18}},\ \bibinfo {pages} {1090} (\bibinfo {year} {2024})}\BibitemShut {NoStop}%
\bibitem [{\citenamefont {Figueiredo}\ \emph {et~al.}(2025)\citenamefont {Figueiredo}, \citenamefont {Schofield}, \citenamefont {Fu}, \citenamefont {Nyman}, \citenamefont {Oulton}, \citenamefont {Terças},\ and\ \citenamefont {Mintert}}]{Figueiredo2025}%
  \BibitemOpen
  \bibfield  {author} {\bibinfo {author} {\bibfnamefont {J.~L.}\ \bibnamefont {Figueiredo}}, \bibinfo {author} {\bibfnamefont {R.~C.}\ \bibnamefont {Schofield}}, \bibinfo {author} {\bibfnamefont {M.}~\bibnamefont {Fu}}, \bibinfo {author} {\bibfnamefont {R.~A.}\ \bibnamefont {Nyman}}, \bibinfo {author} {\bibfnamefont {R.}~\bibnamefont {Oulton}}, \bibinfo {author} {\bibfnamefont {H.}~\bibnamefont {Terças}},\ and\ \bibinfo {author} {\bibfnamefont {F.}~\bibnamefont {Mintert}},\ }\bibfield  {title} {\bibinfo {title} {Photon bose-einstein condensation in semiconductors: A quantum kinetic theory},\ }\href {https://arxiv.org/pdf/2509.05062} {\bibfield  {journal} {\bibinfo  {journal} {arXiv 2509.05062}\ } (\bibinfo {year} {2025})}\BibitemShut {NoStop}%
\bibitem [{\citenamefont {Loirette-Pelous}\ and\ \citenamefont {Greffet}(2023)}]{Pelous2023}%
  \BibitemOpen
  \bibfield  {author} {\bibinfo {author} {\bibfnamefont {A.}~\bibnamefont {Loirette-Pelous}}\ and\ \bibinfo {author} {\bibfnamefont {J.~J.}\ \bibnamefont {Greffet}},\ }\bibfield  {title} {\bibinfo {title} {Photon bose–einstein condensation and lasing in semiconductor cavities},\ }\bibfield  {journal} {\bibinfo  {journal} {Laser and Photonics Reviews}\ }\textbf {\bibinfo {volume} {17}},\ \href {https://doi.org/10.1002/lpor.202300366} {10.1002/lpor.202300366} (\bibinfo {year} {2023})\BibitemShut {NoStop}%
\bibitem [{\citenamefont {Wyborski}\ \emph {et~al.}(2025)\citenamefont {Wyborski}, \citenamefont {Broda}, \citenamefont {Sankowska}, \citenamefont {Czyszanowski}, \citenamefont {Muszalski},\ and\ \citenamefont {Pieczarka}}]{Wyborski2025}%
  \BibitemOpen
  \bibfield  {author} {\bibinfo {author} {\bibfnamefont {P.}~\bibnamefont {Wyborski}}, \bibinfo {author} {\bibfnamefont {A.}~\bibnamefont {Broda}}, \bibinfo {author} {\bibfnamefont {I.}~\bibnamefont {Sankowska}}, \bibinfo {author} {\bibfnamefont {T.}~\bibnamefont {Czyszanowski}}, \bibinfo {author} {\bibfnamefont {J.}~\bibnamefont {Muszalski}},\ and\ \bibinfo {author} {\bibfnamefont {M.}~\bibnamefont {Pieczarka}},\ }\bibfield  {title} {\bibinfo {title} {Thermal equilibrium of light-matter interaction in ingaas/gaas quantum wells},\ }\href {https://doi.org/10.1016/J.JLUMIN.2025.121403} {\bibfield  {journal} {\bibinfo  {journal} {Journal of Luminescence}\ }\textbf {\bibinfo {volume} {286}},\ \bibinfo {pages} {121403} (\bibinfo {year} {2025})}\BibitemShut {NoStop}%
\bibitem [{\citenamefont {Schmitt}(2018)}]{Schmitt2018}%
  \BibitemOpen
  \bibfield  {author} {\bibinfo {author} {\bibfnamefont {J.}~\bibnamefont {Schmitt}},\ }\bibfield  {title} {\bibinfo {title} {Dynamics and correlations of a bose–einstein condensate of photons},\ }\href {https://doi.org/10.1088/1361-6455/AAD409} {\bibfield  {journal} {\bibinfo  {journal} {Journal of Physics B: Atomic, Molecular and Optical Physics}\ }\textbf {\bibinfo {volume} {51}},\ \bibinfo {pages} {173001} (\bibinfo {year} {2018})}\BibitemShut {NoStop}%
\bibitem [{\citenamefont {Kim}\ \emph {et~al.}(2016)\citenamefont {Kim}, \citenamefont {Zhang}, \citenamefont {Wang}, \citenamefont {Fischer}, \citenamefont {Brodbeck}, \citenamefont {Kamp}, \citenamefont {Schneider}, \citenamefont {Höfling},\ and\ \citenamefont {Deng}}]{Kim2016}%
  \BibitemOpen
  \bibfield  {author} {\bibinfo {author} {\bibfnamefont {S.}~\bibnamefont {Kim}}, \bibinfo {author} {\bibfnamefont {B.}~\bibnamefont {Zhang}}, \bibinfo {author} {\bibfnamefont {Z.}~\bibnamefont {Wang}}, \bibinfo {author} {\bibfnamefont {J.}~\bibnamefont {Fischer}}, \bibinfo {author} {\bibfnamefont {S.}~\bibnamefont {Brodbeck}}, \bibinfo {author} {\bibfnamefont {M.}~\bibnamefont {Kamp}}, \bibinfo {author} {\bibfnamefont {C.}~\bibnamefont {Schneider}}, \bibinfo {author} {\bibfnamefont {S.}~\bibnamefont {Höfling}},\ and\ \bibinfo {author} {\bibfnamefont {H.}~\bibnamefont {Deng}},\ }\bibfield  {title} {\bibinfo {title} {Coherent polariton laser},\ }\href {https://doi.org/10.1103/PhysRevX.6.011026} {\bibfield  {journal} {\bibinfo  {journal} {Physical Review X}\ }\textbf {\bibinfo {volume} {6}},\ \bibinfo {pages} {011026} (\bibinfo {year} {2016})}\BibitemShut {NoStop}%
\bibitem [{\citenamefont {Drechsler}\ \emph {et~al.}(2022)\citenamefont {Drechsler}, \citenamefont {Lohof},\ and\ \citenamefont {Gies}}]{Drechsler2022}%
  \BibitemOpen
  \bibfield  {author} {\bibinfo {author} {\bibfnamefont {M.}~\bibnamefont {Drechsler}}, \bibinfo {author} {\bibfnamefont {F.}~\bibnamefont {Lohof}},\ and\ \bibinfo {author} {\bibfnamefont {C.}~\bibnamefont {Gies}},\ }\bibfield  {title} {\bibinfo {title} {Revisiting the siegert relation for the partially coherent regime of nanolasers},\ }\href {https://doi.org/10.1063/5.0094698} {\bibfield  {journal} {\bibinfo  {journal} {Applied Physics Letters}\ }\textbf {\bibinfo {volume} {120}},\ \bibinfo {pages} {221104} (\bibinfo {year} {2022})}\BibitemShut {NoStop}%
\bibitem [{\citenamefont {Ulrich}\ \emph {et~al.}(2007)\citenamefont {Ulrich}, \citenamefont {Gies}, \citenamefont {Ates}, \citenamefont {Wiersig}, \citenamefont {Reitzenstein}, \citenamefont {Hofmann}, \citenamefont {Löffler}, \citenamefont {Forchel}, \citenamefont {Jahnke},\ and\ \citenamefont {Michler}}]{Ulrich2007}%
  \BibitemOpen
  \bibfield  {author} {\bibinfo {author} {\bibfnamefont {S.~M.}\ \bibnamefont {Ulrich}}, \bibinfo {author} {\bibfnamefont {C.}~\bibnamefont {Gies}}, \bibinfo {author} {\bibfnamefont {S.}~\bibnamefont {Ates}}, \bibinfo {author} {\bibfnamefont {J.}~\bibnamefont {Wiersig}}, \bibinfo {author} {\bibfnamefont {S.}~\bibnamefont {Reitzenstein}}, \bibinfo {author} {\bibfnamefont {C.}~\bibnamefont {Hofmann}}, \bibinfo {author} {\bibfnamefont {A.}~\bibnamefont {Löffler}}, \bibinfo {author} {\bibfnamefont {A.}~\bibnamefont {Forchel}}, \bibinfo {author} {\bibfnamefont {F.}~\bibnamefont {Jahnke}},\ and\ \bibinfo {author} {\bibfnamefont {P.}~\bibnamefont {Michler}},\ }\bibfield  {title} {\bibinfo {title} {Photon statistics of semiconductor microcavity lasers},\ }\href {https://doi.org/10.1103/PhysRevLett.98.043906} {\bibfield  {journal} {\bibinfo  {journal} {Physical Review Letters}\ }\textbf {\bibinfo {volume} {98}},\ \bibinfo {pages} {043906} (\bibinfo {year} {2007})}\BibitemShut {NoStop}%
\bibitem [{\citenamefont {Tang}\ \emph {et~al.}(2024)\citenamefont {Tang}, \citenamefont {Dhar}, \citenamefont {Oulton}, \citenamefont {Nyman},\ and\ \citenamefont {Mintert}}]{Tang2023a}%
  \BibitemOpen
  \bibfield  {author} {\bibinfo {author} {\bibfnamefont {Y.}~\bibnamefont {Tang}}, \bibinfo {author} {\bibfnamefont {H.~S.}\ \bibnamefont {Dhar}}, \bibinfo {author} {\bibfnamefont {R.~F.}\ \bibnamefont {Oulton}}, \bibinfo {author} {\bibfnamefont {R.~A.}\ \bibnamefont {Nyman}},\ and\ \bibinfo {author} {\bibfnamefont {F.}~\bibnamefont {Mintert}},\ }\bibfield  {title} {\bibinfo {title} {Breakdown of temporal coherence in photon condensates},\ }\href {https://doi.org/10.1103/PhysRevLett.132.173601} {\bibfield  {journal} {\bibinfo  {journal} {Physical Review Letters}\ }\textbf {\bibinfo {volume} {132}},\ \bibinfo {pages} {173601} (\bibinfo {year} {2024})}\BibitemShut {NoStop}%
\bibitem [{\citenamefont {Wagner}\ \emph {et~al.}(2021)\citenamefont {Wagner}, \citenamefont {Wagner}, \citenamefont {Schiffers}, \citenamefont {Willomitzer}, \citenamefont {Cossairt}, \citenamefont {Cossairt}, \citenamefont {Velten},\ and\ \citenamefont {Velten}}]{Wagner2021}%
  \BibitemOpen
  \bibfield  {author} {\bibinfo {author} {\bibfnamefont {F.}~\bibnamefont {Wagner}}, \bibinfo {author} {\bibfnamefont {F.}~\bibnamefont {Wagner}}, \bibinfo {author} {\bibfnamefont {F.}~\bibnamefont {Schiffers}}, \bibinfo {author} {\bibfnamefont {F.}~\bibnamefont {Willomitzer}}, \bibinfo {author} {\bibfnamefont {O.}~\bibnamefont {Cossairt}}, \bibinfo {author} {\bibfnamefont {O.}~\bibnamefont {Cossairt}}, \bibinfo {author} {\bibfnamefont {A.}~\bibnamefont {Velten}},\ and\ \bibinfo {author} {\bibfnamefont {A.}~\bibnamefont {Velten}},\ }\bibfield  {title} {\bibinfo {title} {Intensity interferometry-based 3d imaging},\ }\href {https://doi.org/10.1364/OE.412688} {\bibfield  {journal} {\bibinfo  {journal} {Optics Express, Vol. 29, Issue 4, pp. 4733-4745}\ }\textbf {\bibinfo {volume} {29}},\ \bibinfo {pages} {4733} (\bibinfo {year} {2021})}\BibitemShut {NoStop}%
\end{thebibliography}%

\end{document}